\documentclass[preprint,aps,showpacs,nofootinbib,superscriptaddress,preprintnumbers,epsf,psf]{revtex4}
\usepackage[english]{babel}
\usepackage{pstricks}
\usepackage[dvips]{graphicx}
\usepackage{epsf}
\usepackage{amsmath}
\usepackage{amsfonts}
\usepackage{amssymb}

\usepackage{color}
\usepackage{dcolumn}
\usepackage{bm}
\everymath{\displaystyle}
\begin{document}

\title{Femto-cyclones and hyperon  polarization in Heavy-Ion Collisions }

\author{\firstname{Mircea}~\surname{Baznat}}
\email{baznat@theor.jinr.ru}
\author{\firstname{Konstantin}~\surname{Gudima}}
\email{gudima@theor.jinr.ru} \affiliation{Joint Institute for Nuclear Research, 141980 Dubna (Moscow region),
Russia} \affiliation{Institute of Applied Physics, Academy of Sciences
of Moldova, MD-2028 Kishinev, Moldova }
\author{\firstname{Alexander}~\surname{Sorin}}
\email{sorin@theor.jinr.ru}
\author{\firstname{Oleg}~\surname{Teryaev}}
\email{teryaev@theor.jinr.ru} \affiliation{Joint Institute for Nuclear Research, 141980 Dubna (Moscow region), Russia}
\affiliation{National Research Nuclear University MEPhI (Moscow Engineering Physics Institute),
Kashirskoe Shosse 31, 115409 Moscow, Russia}

\date{\today}

\begin {abstract}
We study the structure of vorticity and hydrodynamic helicity fields in peripheral 
heavy-ion collisions using the kinetic Quark-Gluon String Model. 
The angular momentum conservation within this model holds with a good accuracy.
We observe the formation of specific toroidal structures of vorticity field (vortex sheets).
Their existence is mirrored in the polarization of hyperons of the percent 
order.

\end{abstract}


\maketitle

\section{Introduction}

The local violation \cite{Fukushima:2008xe} of discrete symmetries
in strongly interacting QCD matter is now under intensive
theoretical and experimental investigations. The renowned Chiral
Magnetic Effect (CME) uses the (C)P-violating (electro)magnetic
field emerging in heavy ion collisions in order to probe the
(C)P-odd effects in QCD matter.

There is an interesting counterpart of this effect, Chiral
Vortical Effect (CVE)\cite{Kharzeev:2007tn} due to coupling to
P-odd medium vorticity leading to the induced 
electromagnetic and all
conserved-charge currents \cite{Rogachevsky:2010ys}, in particular the 
baryonic one.

Another important P-odd observable is the baryon polarizatiion. 
The mechanism analogous to CVE (known as axial vortical effect, see \cite{Kalaydzhyan:2014bfa} and references therein) 
leads to induced axial 
current of strange quarks which may be converted to polarization 
of $\Lambda$-hyperons \cite{Rogachevsky:2010ys,Gao:2012ix,Baznat:2013zx}. 
Another mechanism of this polarization is 
provided  by so-called thermal vorticity in the hydrodynamical approach \cite{Becattini:2013vja}.   

The zeroth component of axial current and correspondent axial charge are related to hydrodynamical helicity 
$$H \equiv \int d V  (\vec v \cdot \vec w),$$ being the projection of velocity $\vec v$ to vorticity $\vec w= curl \vec v$.
This quantity is 
manifesting the recently discovered \cite{Baznat:2013zx} and confirmed \cite{Teryaev:2014qia} phenomenon of the separation, 
i.e. its mirror behavior with the same magnitudes but different signs in the half-spaces separated by the reaction plane.

The noncentral heavy ion collisions could naturally generate a rotation
(global or local, both related to vorticity)
with an angular velocity normal to the reaction plane,
which is their generic qualitative feature. It is naturally to expect that angular momentum conservation 
plays an essential role in the defining the quantitative properties of vortical effects. At the same time,
it remains to be studied to which extent the particles carrying the main part of angular momentum participate
in the collisions.

In the current paper we address these problem by performing the extensive numerical simulations. 
We explore the distribution of angular momentum and find that the role of participant nucleons 
is relatively small, albeit noticeable. 
We study in some detail the structure of vorticity field and compare different approaches to polarization calculation.
We observed the peculiar toroidal "tire-like" structure manifesting themselves in the polarization of hyperons.
We will explore different approaches to polarization calculation which will lead to qualitatively similar results.

\section{Angular momentum conservation in the kinetic model}
 
The natural source of the P-odd observables in heavy-ion collisions is the 
pseudovector of angular momentum. The question immediately emerges whether it is 
conserved in the course of evolution governed by Quark-Gluon String Model (QGSM) 
\cite{toneev90}.
To check this we calculated the angular momentum at various time moments of collision 
taking into account both the contributions of participants and spectators.
We consider the $Au+Au$ collisions with $b = 8~fm$ at $\sqrt{s} = 5\,GeV/u$ typical for future NICA collider. 
We observed (see Fig.1) that the participants carry about $20\%$ of angular momentum and that the total angular momentum
of participants and spectators is conserved with a rather good accuracy. 

 \begin{figure}[h!]
\includegraphics
[height=0.5\textheight]
{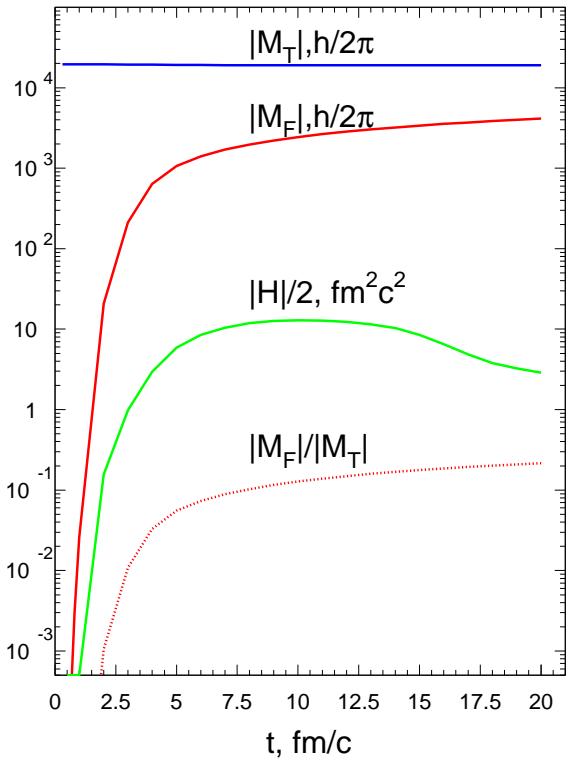} 
\caption{The time dependence of the total ($M_T$) and fireball ($M_F$) angular momenta in Planck constant units and that of hydrodynamical helicity ($H$).}
\end{figure}

One may conclude that the angular momentum is under the good control in QGSM model. 

We also studied correlations of angular momentum  and  
the hydrodynamic helicity and observed that these quantities vary in accordance with each other. 
In order to perform this comparison and further calculations, velocity, vorticity and helicity were determined following the 
earlier suggested \cite{Baznat:2013zx} procedure when the respective quantities were 
properly averaged over events and particles within the three-dimensional cells providing the 
transition from the kinetic to hydrodynamic description. 

Let us pass to the corresponding results.

\section{Large-scale structures of vorticity fields}

We start our studies with the qualitative structure of velocity and vorticity fields.

The general structure of velocity field follows the "little bang" pattern
which may be quantified by the  velocity 
dependence allowing to extract the "little Hubble" constant. 
We calculated the dependence of average cell velocity on the transverse distance  $\rho=\sqrt{x^2 +  
y^2}$ and found (see Fig.2) that 
it is consistent with the linear "Hubble" law
\begin{equation}
<v/c> = v_0/c + H\rho.
\label{H}
\end{equation}
The Hubble constant $H$ is measured in the units $10^{-22}\,sec\,=\,30\,fm/c$ and changes in the range
 $$H\,=\,0.024 \,\div \,  0.028 \, (fm/c)^{-1}.$$
It corresponds to the "little Universe" lifetime of about $40\, fm/c$ which is only twice larger than the
 collision time. 

\begin{figure}[h!]
\includegraphics[angle=-0,width=0.5\textheight]{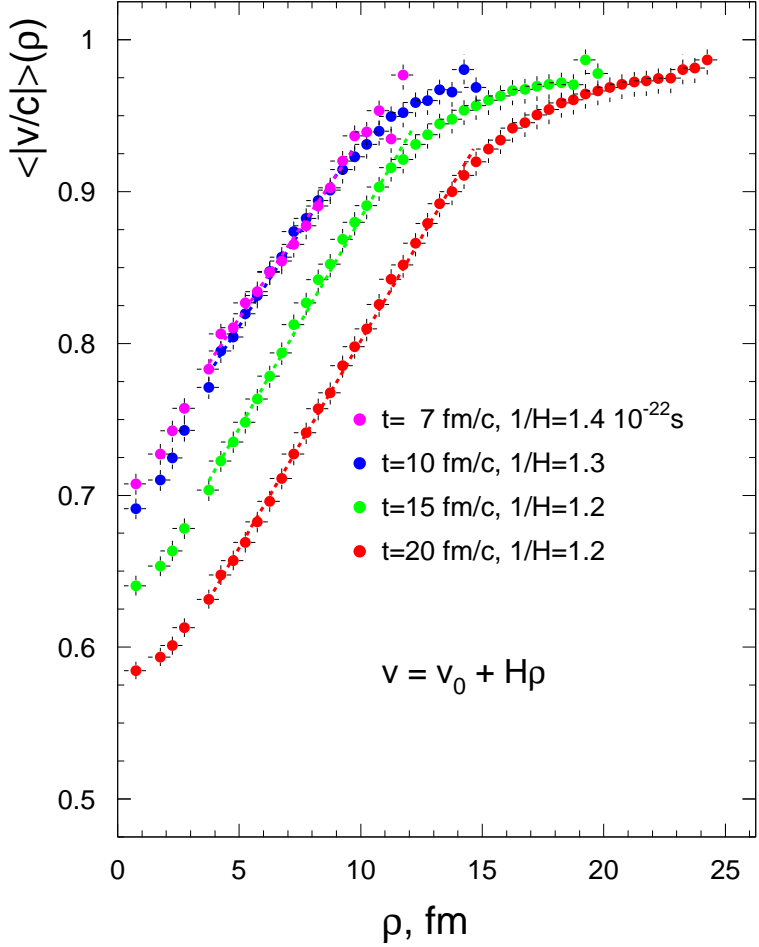} 
\caption{The cell velocity dependence on the transverse distance.}
\end{figure} 

Our key observation is that while velocity field represent the "little bang" 
picture, vorticity field form the relatively thin
toroidal "tire-like" structures (Fig. 3),
\begin{figure}[h!]
\includegraphics[height=0.8\textwidth]{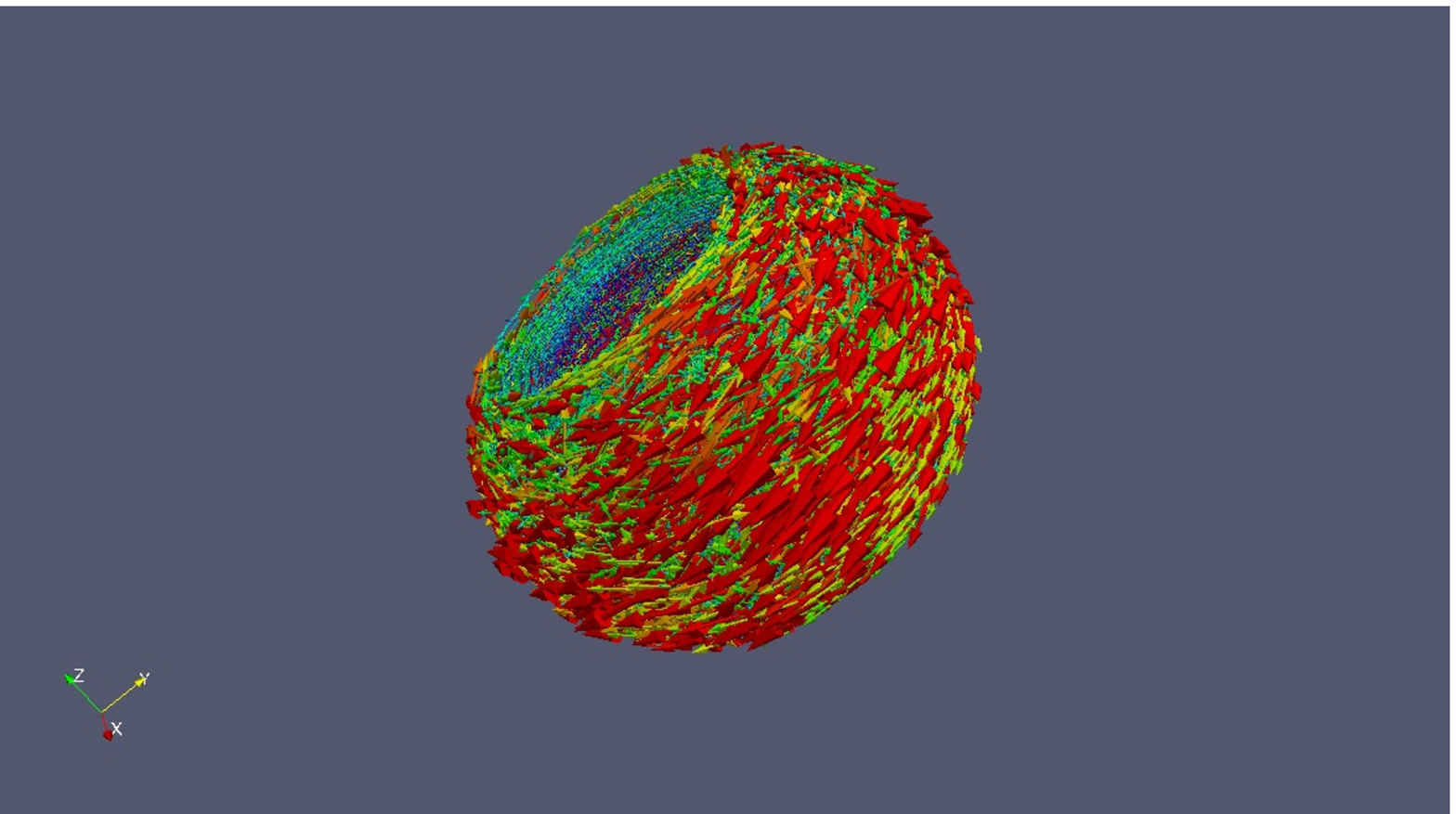}
\caption{The vortex sheet.}
\end{figure}
which emerge in the layer (where velocity field changes rapidly) separating the "core" and "corona" regions \cite{Aichelin,St.& Bl.} and 
form the sort of vortex sheet.
  
The interesting property of these structures is that, while emerging due to angular momentum pseudovector $\vec M$
in the non-central collisions they do not "remember" the production plane and  possess the cylindrical symmetry w.r.t. collisions axis $z$. 
This may be observed (Fig.4) by considering the vortex sheet in the case of the particular direction of $\vec M$ along the $y$ axis.

 \begin{figure}[h!]
 \includegraphics[angle=-0,height=0.5\textheight]{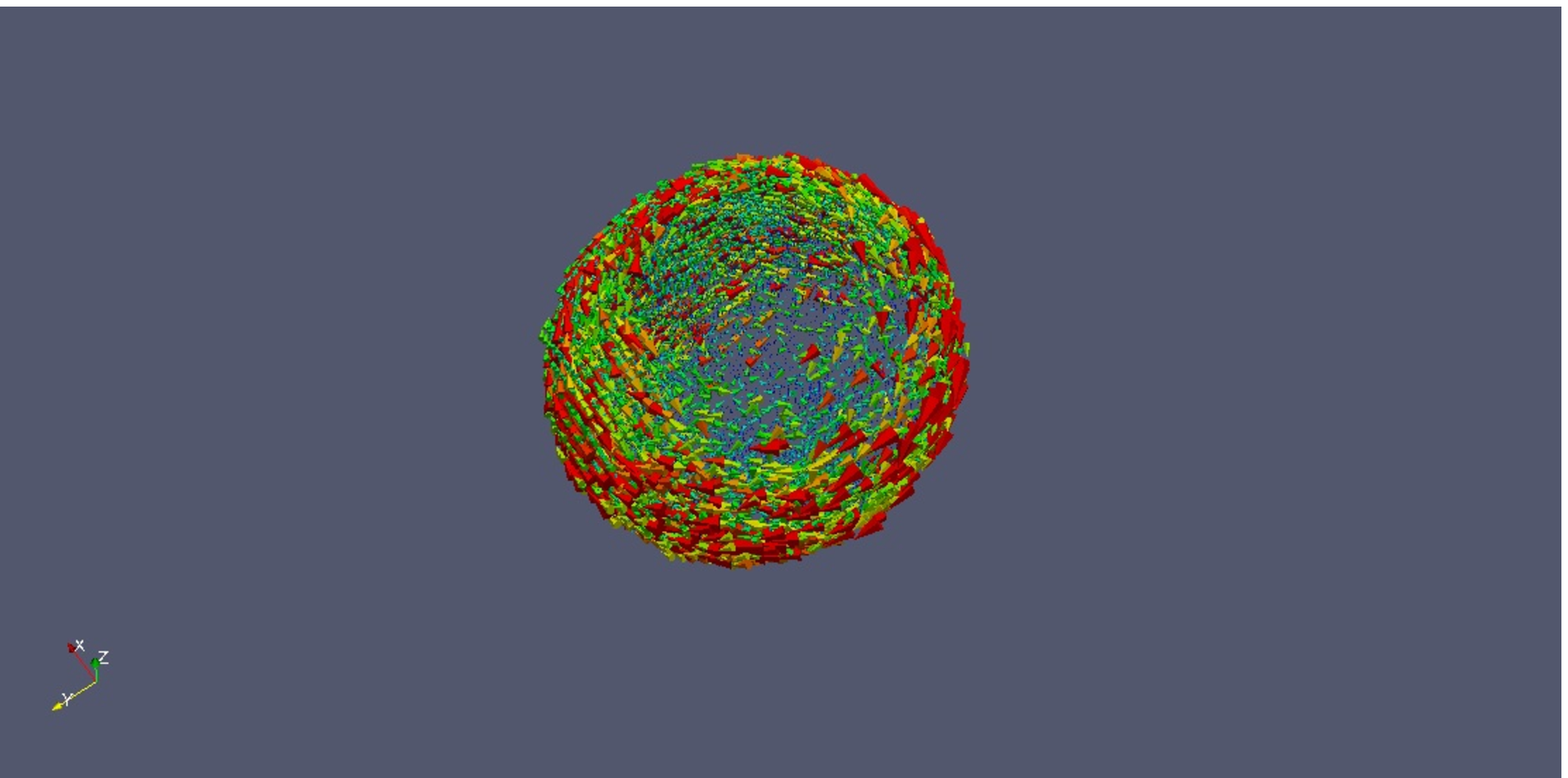}
\caption{The vortex sheet for the particular direction of angular momentum.}
\end{figure}

Such behaviour may resemble cyclones appearing at femtoscopic scale.

Let us now discuss the observable signatures of these nice structures.

\section{Hyperon polarization}

We consider hyperon polarization as the observable related  to vorticity and helicity \cite{Rogachevsky:2010ys}. 
We shall concentrate mostly on $\Lambda$ hyperon production, which has some advantages: they are produced in large numbers, 
their polarization may be easily determined in their weak decays, and their spin is carried by strange quark.

We compare the two rather distinct methods of determining the hyperon polarization.
The first corresponds to its earlier suggested \cite{Rogachevsky:2010ys} and explored \cite{Baznat:2013zx} relation to the induced axial current while the second one 
follows the procedure  based on the thermal vorticity \cite{Becattini:2013vja}.

The first method is based on the calculation of strange axial charge

\begin{eqnarray}
\label{q5s}
Q_5^s=\frac{N_c}{{2 \pi^2}} \int d^3 x \mu^2(x) \gamma^2 \epsilon^{i j k}u_{i} \partial_{j}u_ k \nonumber \\ = 
\frac{<\mu^2 \gamma^2> N_c H}{2 \pi^2}.
\end{eqnarray}
In \cite{Baznat:2013zx} we used the latter equality exploring the mean-value theorem, while here the spatial variation  
of strange chemical potential $\mu$ is taken into account. To do so, the description of kinetic distribution functions by the correspondent equilibrium 
equation was performed, providing the matching of kinetic and thermodynamical descriptions. 
As a result, the time dependence of the distribution of strange chemical potential takes the form represented at Fig. 5. 
 \begin{figure}[h!]
 \hspace{20mm}
\includegraphics[angle=-0,height=0.7\textheight]{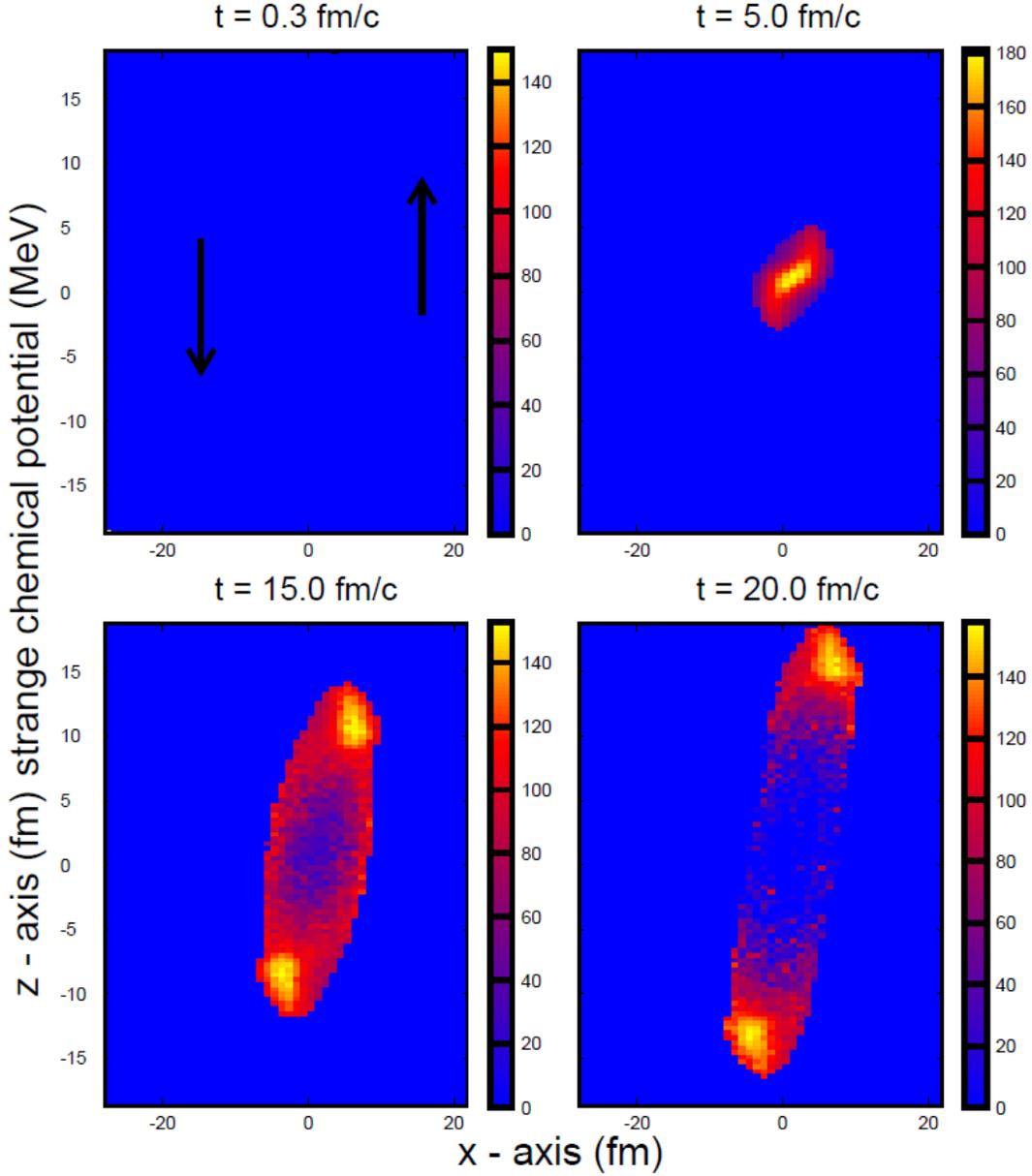}
 \caption{The time dependence of strange chemical potential.}
\end{figure}
The average polarization can be estimated by dividing $Q_5^s$ (\ref{q5s}) by the number of $\Lambda$'s, assuming that the pseudovector of axial current is proportional to the pseudovector of polarization, $Q_5^s \,\sim\,<\Pi_0^{\Lambda,lab}>$. Selecting the axial charge related to the particles in the definite rapidity or transverse momentum interval, the respective dependence of polarization may be also obtained.

As the axial charge should be related to the zeroth component of hyperon polarization in laboratory frame $\Pi_0^{lab}$, the 
transformation to hyperon rest frame should be performed. Taking into account that polarization pseudovector 
should be directed along $y$ axis (as it has to be collinear to $\vec M$ pseudovector), one get
\begin{equation}
\label{Pi}
\Pi_0^{\Lambda,lab} =  \frac{\Pi_0^{\Lambda}\, p_y }{M_{\Lambda}},
\end{equation} 
so that the rest frame polarization can be obtained as 
\begin{equation}
\label{ratio}
<\Pi_0^{\Lambda}> \,= \,Q_5^s <\frac{M_{\Lambda}}{N_{\Lambda} \,p_y }>.
\end{equation} 
The possible violation of positivity ($|\Pi_{\Lambda}| \geq 1$) or even the divergence is due to the fact, that
hyperons with zero $y$ component of the momentum should not have the zeroth component of polarization and therefore should not 
contribute to $Q_5^s$. To avoid this problem one may instead attribute the factor $p_y/M$ to each hyperon in the denominator 
of (\ref{ratio}). Nevertheless the comparison (see  Fig. 6) of various approaches shows  the similar scale and rapidity dependence of polarization. 

 \begin{figure}[h!]
\label{figpolq5} 
\hspace{-20mm}
\includegraphics[angle=-0,width=1\columnwidth]{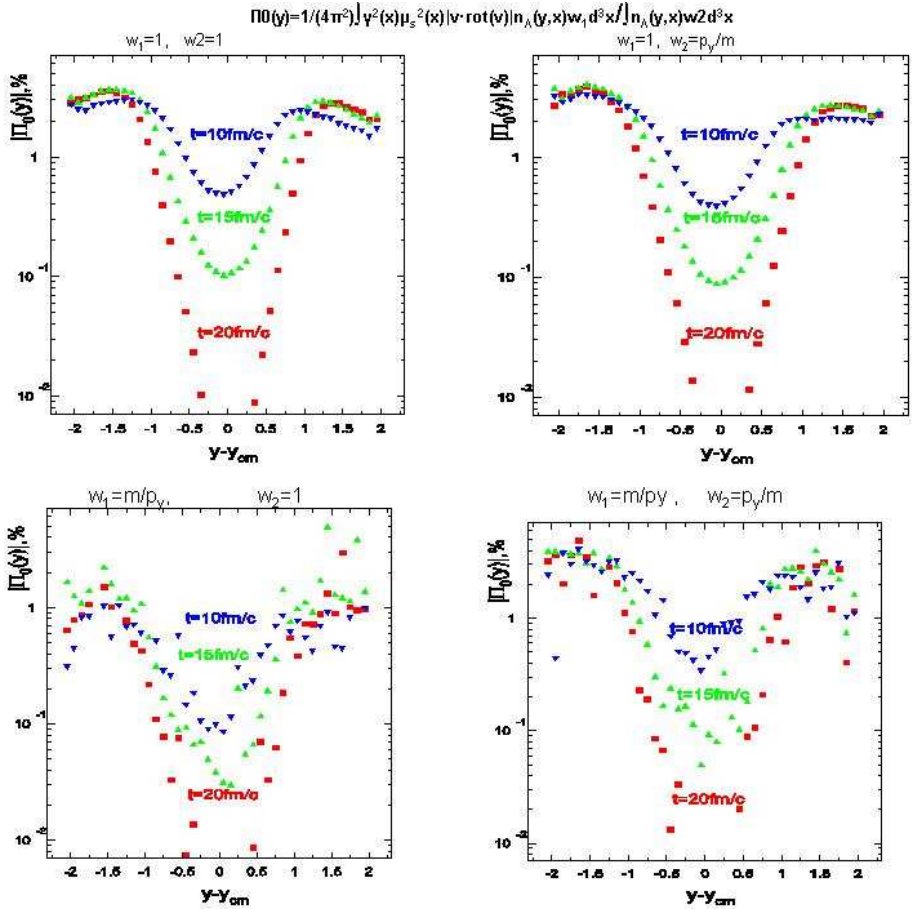}
 \caption{The rapidity dependence of polarization in helicity-based approach.}
\end{figure}

Another approach to polarization is based on the so-called thermal vorticity \cite{Becattini:2013vja}.
To provide the comparison we calculated (see Fig. 7) the thermal vorticity field and respective polarization. 
  \begin{figure}[h!]
 \hspace{-20mm}
\includegraphics[angle=-0,height=0.7\textheight]{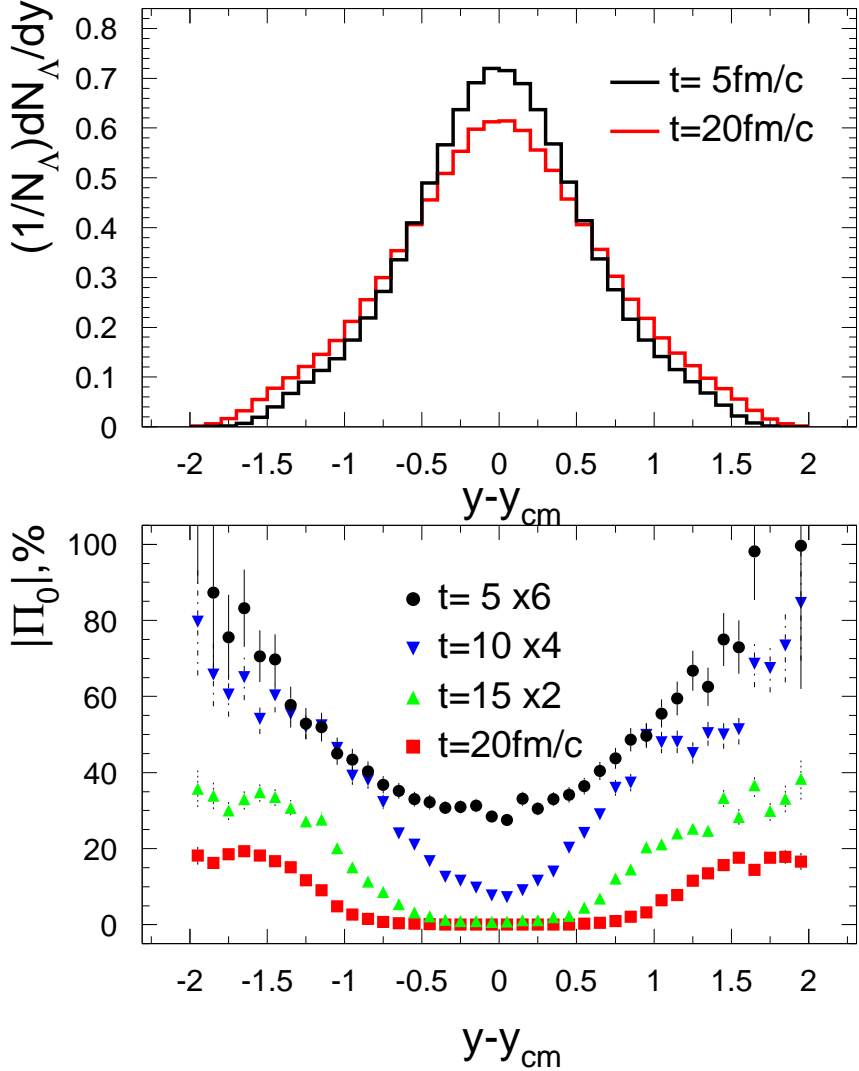}
 \caption{The rapidity dependence of the multiplicity and polarization in the thermal vorticity-based approach. The magnitudes of polarization at different times are rescaled by the identified factors.}
\end{figure}
While a scale of the polarization in the thermal vorticity-based approach is several times larger, its rapidity dependence, surprisingly enough,  appear to be similar in these rather distinct approaches.

\section{Conclusions and Outlook}

We investigated vorticity and hydrodynamical helicity in noncentral heavy-ion collisions in the framework of the kinetic Quark-Gluon String Model.
We have confirmed our earlier observation that the vorticity field is predominantly localized in a relatively thin layer ($2\div 3~fm$) on the boundary between the participant and spectator nucleons and observed that it is forming the specific toroidal structures,
which might be considered as  vortex sheets with the unexpected cylindrical symmetry. They look as  cyclones appearing at femtoscopic scale.   

The vorticity and helicity fields are manifested in the $\Lambda$ hyperons polarization. We performed its detailed calculations
including the simulations of the strange chemical potential. We found that the polarization magnitude may reach a percent level. 
The comparison with the very different approach exploring the thermal vorticity leads to qualitatively similar results, although the polarization scale is several times larger.

{}~

{\bf Acknowledgements}

{

 This work was
supported in part by the Russian Foundation for Basic Research,
Grant No. 14-01-00647.


\end{document}